%% file: dred_journal.tex
\documentclass{IEEEtran}
\usepackage[latin9]{inputenc}
\usepackage{amsmath}
\usepackage{amssymb}
\usepackage{graphicx}

\makeatletter
\usepackage[pdftex,pdftitle={DRED: Deep REDundancy Coding of Speech Using a Rate-Distortion-Optimized Variational Autoencoder},pdfauthor={Jean-Marc Valin, Jan Buethe, Ahmed Mustafa, Michael Klingbeil}]{hyperref}
\usepackage{balance}

\usepackage{tikz}
\usetikzlibrary{shapes.geometric, shapes.arrows, arrows, backgrounds, math}
\tikzstyle{arrow}=[draw, -latex]

\makeatother

\begin{document}
\title{DRED: Deep REDundancy Coding of Speech Using a Rate-Distortion-Optimized
Variational Autoencoder}
\author{Jean-Marc Valin, \IEEEmembership{Member, IEEE}, Jan Büthe, \IEEEmembership{Member, IEEE},
Ahmed Mustafa, Michael Klingbeil\thanks{Jean-Marc Valin and Jan B\"{u}the are with the Xiph.Org Foundation, but the work was conducted while with Amazon Web Services (e-mail: \texttt{jmvalin@jmvalin.ca}, \texttt{jan.buethe@ieee.org})}
\thanks{Ahmed Mustafa and Michael Klingbeil are with Amazon Web Services (e-mail: \texttt{ahdmust@amazon.com}, \texttt{klingm@amazon.com}).}}

\maketitle
\maketitle 
\begin{abstract}
Despite recent advancements in packet loss concealment (PLC) using
deep learning techniques, packet loss remains a significant challenge
in real-time speech communication. Redundancy has been used in the
past to recover the missing information during losses. However, conventional
redundancy techniques are limited in the maximum loss duration they
can cover and are often unsuitable for burst packet loss. We propose
a new approach based on a rate-distortion-optimized variational autoencoder
(RDO-VAE), allowing us to optimize a deep speech compression algorithm
for the task of encoding large amounts of redundancy at very low bitrate.
The proposed Deep REDundancy (DRED) algorithm can transmit up to 50x
redundancy using less than 32 kb/s. Results show that DRED outperforms
the existing Opus codec redundancy. We also demonstrate its benefits
when operating in the context of WebRTC.
\end{abstract}

\begin{IEEEkeywords}neural speech coding, audio redundancy, variational
autoencoder\end{IEEEkeywords} 

\section{Introduction}

Recent advancements in Deep Packet Loss Concealment (PLC)~\cite{plc_challenge}
have demonstrated significant improvements in quality, but also the
fundamental limitations of PLC, especially when it comes to intelligibility.
For long burst loss, information about missing syllables/words cannot
-- and should not -- be predicted. In parallel, neural speech codecs~\cite{kleijn2018wavenet,klejsa2019,valin2019lpcnetcodec}
have also demonstrated their ability to efficiently transmit speech
with sufficient quality at very low bitrates and we believe they can
play a part in improving how we address packet loss.

One way to improve robustness to lost packets is to transmit redundant
information. Some methods, such as RED~\cite{rfc2198} are generic,
whereas the Opus~\cite{rfc6716} low-bitrate redundancy (LBRR) is
specific to that codec. While techniques like LBRR make it possible
to transmit redundancy at a somewhat lower bitrate than normal, they
cannot scale to large amounts of redundancy without causing impractically
large overheads in bitrates.

\begin{figure}

\begin{centering}
\includegraphics[width=1\columnwidth]{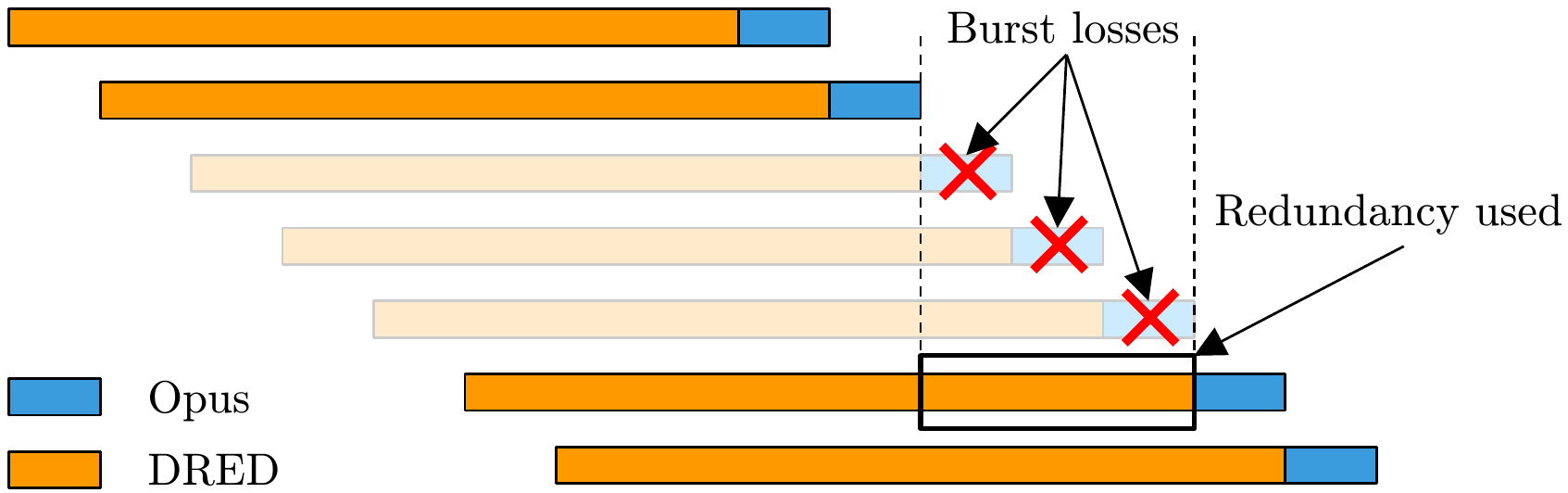}
\par\end{centering}
\caption{Illustrating deep redundancy for the case of three consecutive losses.
The first packet to arrive after the loss includes enough redundant
information to reconstruct the missing audio. \protect\label{fig:Illustrating-redundancy}}

\end{figure}

In this work, we propose adapting recent neural speech coding techniques
to the problem of transmitting redundant audio information. The architecture
of our Deep REDundancy (DRED) mechanism (Fig.~\ref{fig:Illustrating-redundancy})
takes advantage of a continuously-operating recurrent encoder with
a decoder running backward in time (Section~\ref{sec:Overview})
to satisfy the unique constraints of redundancy coding. At a lower
level, a rate-distortion-optimized variational autoencoder (RDO-VAE)
produces a Laplace-distributed latent space that can be quantized
efficiently (Section~\ref{sec:RDO-VAE}). We demonstrate (Section~\ref{sec:Experiments-Results})
that DRED can be used to guard against burst losses up to 1 second
(50x redundancy) using less than 32 kb/s, a significant improvement
over the Opus LBRR. Results also show that these improvements translate
into significant quality improvements in the context of a WebRTC implementation. 

This paper is an extended version of~\cite{valin2023dred}, including
more a detailed description and additional experiments. The algorithm
has also been improved with a more efficient encoder-decoder architecture,
the addition of rate control to the initial states, and the replacement
of the LPCNet vocoder with a new auto-regressive vocoder~\cite{valin2024fargan},
reducing synthesis complexity by a factor of~5. The resulting DRED
algorithm is deployed in Opus~1.5, for which we provide an updated
evaluation.

\section{Deep Redundancy (DRED) Overview}

\label{sec:Overview}

\begin{figure*}

\begin{centering}
\includegraphics[width=0.75\linewidth]{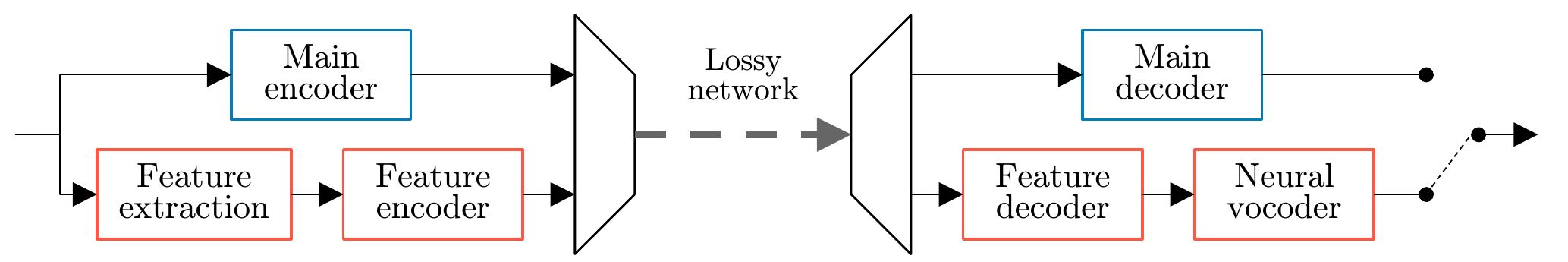}
\par\end{centering}
\caption{High-level overview of a communication system using deep redundancy.\protect\label{fig:High-level-overview}}

\end{figure*}

Most speech codecs in use today encode audio in 20\nobreakdash-ms
frames, with each frame typically being sent in a separate packet
over the Internet. When any packet is lost, the corresponding audio
is lost and has to be filled by a PLC algorithm. The Opus LBRR option
makes it possible for packet number $n$ to include the contents of
both frames $n$ and $n-1$, with the latter being encoded at a slightly
lower bitrate. Effectively, packets each contains 40\nobreakdash-ms
of audio despite being sent at a 20\nobreakdash-ms interval. When
LBRR is enabled, a single packet loss does not cause any audio frame
to be completely lost, and this can improve the quality in difficult
network conditions. Unfortunately, losses are rarely uniformly distributed,
and LBRR has limited impact on long loss bursts. While more frames
could be coded as part of each packet, it would cause the bitrate
to go up significantly. For that reason, we propose an efficient deep
speech coding technique that makes it possible to include a large
amount of redundancy without a large increase in bitrate. The redundancy
is encoded separately, but multiplexed with the regular packets during
transmission (Fig.~\ref{fig:High-level-overview}) in a way that
preserves compatibility with the original Opus specification. On the
receiver side, it is only used when packets are lost. Although one
could design a complete communication system using only the redundancy
path, the ``legacy'' Opus coding still has higher no-loss quality
and better handling of non-speech signals. Just like LBRR, DRED is
only targeted at speech.

The proposed deep redundancy coding scheme exploits the fact that
neural vocoders can generate high-quality speech from low-dimensional
acoustic features. For the present application, these are chosen to
be 18-Bark frequency cepstral coefficients (BFCC), a pitch period
estimated according to~\cite{subramani2023}, and a pitch correlation
estimated on the low-passed speech. These features are computed on
20-ms overlapping windows at a 10-ms interval, making it easy to interface
with speech communication systems that typically operate with multiples
of 10~ms.

The signal-level architecture for the proposed redundancy coding is
derived from our previous work on packet loss concealment, where a
vocoder is used to fill in the missing frames using acoustic features
produced by a predictor (Section~4.3 of~~\cite{valin2022plc}).
In this work, we replace the acoustic feature predictor by an encoder
and decoder that transmit a lossy approximation of the ground-truth
features. While~\cite{yang2023neural} proposes coding a prediction
residual, we instead project the features onto a latent space to further
improve coding efficiency. Although we only discuss redundant audio
coding here, our architecture makes it easy to integrate redundancy
coding with PLC. 

\subsection{Constraints and assumptions}

Since the purpose of this work is to improve robustness to packet
loss, an obvious constraint is to avoid any prediction across different
packets. That being said, within each packet, any amount of prediction
is allowed since we assume that a packet either arrives uncorrupted,
or does not arrive at all. Additionally, since the same frame information
is encoded in multiple packets, we do not wish to re-encode each packet
\emph{from scratch}, but rather have a continuously-running encoder
from which we extract overlapping encoded frames. On the decoder side,
since short losses are more likely than very long ones, it is desirable
to be able to decode only the last few frames of speech without having
to decode the entire packet. 

To maximize efficiency, we can take advantage of (variable-length)
entropy coding. Even if a constant bitrate was ultimately desired,
that could easily be achieved by varying the duration of the redundancy.
We can also take advantage of variable encoding quality as a function
of the timestamp within the redundant payload. Since more recent redundancy
is expected to be used more often, it deserves to be coded at a higher
quality.

Although there are many different types of neural vocoders, we propose
to use an auto-regressive vocoder, as it allows for seamless transitions
between regular coded audio and low-bitrate redundant audio without
the use of cross-fading. Although our previous work used LPCNet~\cite{valin2019lpcnet},
we use the newer Framewise Autoregressive GAN (FARGAN)~\cite{valin2024fargan}
vocoder, which has better quality than LPCNet, and about 1/5th of
the complexity. FARGAN's efficiency is achieved by combining framewise
generation~\cite{mustafa2023framewise} with a pitch-based autoregressive
component~\cite{morrison2021chunked}. It generates speech one 2.5-ms
sub-frame at a time, based on the previous sub-frame as well as an
explicit long-term pitch prediction based on the previously synthesized
speech. Unlike many other autoregressive vocoders, it is trained in
closed loop (unrolling the network) without the use of teacher forcing.
FARGAN is conditioned on the same 10-ms frame features as LPCNet. 

\subsection{Proposed architecture}

\begin{figure}
\begin{centering}
\includegraphics[width=1\columnwidth]{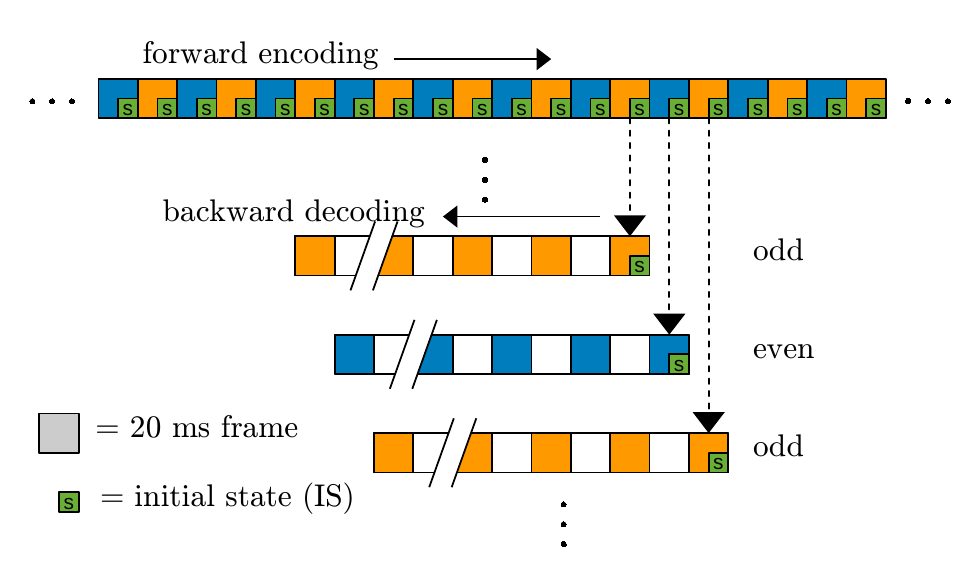}\vspace{0.01cm}
\par\end{centering}
\caption{Overview of the encoding and decoding process. For each 20\protect\nobreakdash-ms
frame, the encoder processes two 10\protect\nobreakdash-ms feature
vectors and produces an encoded latent vector (shown in orange or
blue), as well as an initial state (IS). Although latent vectors are
produced every 20~ms, they each contain sufficient information to
reconstruct 40~ms of audio. The encoded stream is \emph{split} into
overlapping redundancy packets. Each packet to be sent contains a
single IS (for the latest frame), as well as half of the latent vectors
(even or odd) spanning the desired redundancy duration.\protect\label{fig:encoder-decoder}}
\end{figure}

There are generally two methods for improving coding efficiency: prediction
and transforms. The proposed algorithm leverages both methods. In
the context of neural coding, grouping input feature vectors together
enables the encoder to infer an efficient non-linear transform of
its input. For prediction, we use a recurrent neural network (RNN)
architecture, but to achieve the computational goals listed above,
we make the encoder RNN run forward in a continuous manner, while
making the decoder RNN run backward in time from the most recent packet
encoded. Since the process is analogous to delta coding, the encoder
also codes an \emph{initial state} (IS) to ensure that the decoder
achieves sufficient quality on the first (most recent) packet. The
IS vector contains the most recent information and is used to initialize
the decoder. Although the encoder needs to produce such an IS on every
frame, only one (the most recent) is included in each redundancy packet. 

Even though our network operates on 20\nobreakdash-ms frames, the
underlying 20\nobreakdash-dimensional acoustic feature vectors are
computed on a 10\nobreakdash-ms interval. For that reason we group
feature vectors in pairs -- equivalent to a 20\nobreakdash-ms non-linear
transform. To further increase the effective transform size while
still producing a redundancy packet every 20~ms, we use an output
stride. The process is illustrated for a stride of~2 in Fig.~\ref{fig:encoder-decoder}
-- resulting in each output vector representing 40~ms of speech
-- but it can easily scale to larger strides. 

\section{Rate-Distortion-Optimized VAE}

\label{sec:RDO-VAE}

As stated above, our goal is to compress each redundancy packet as
efficiently as possible. Although VQ-VAE~\cite{van2017neural} has
been a popular choice for deep speech coding~\cite{zeghidour2021,casebeer2021},
in this work we avoid its large fixed-size codebooks and investigate
other variational auto-encoders (VAE)~\cite{kingma2013}. Our approach
is instead inspired from recent work in VAE-based image coding~\cite{guo2021,balle2021}
combining scalar quantization with entropy coding.

\begin{figure}
\begin{centering}
\includegraphics[width=1\columnwidth]{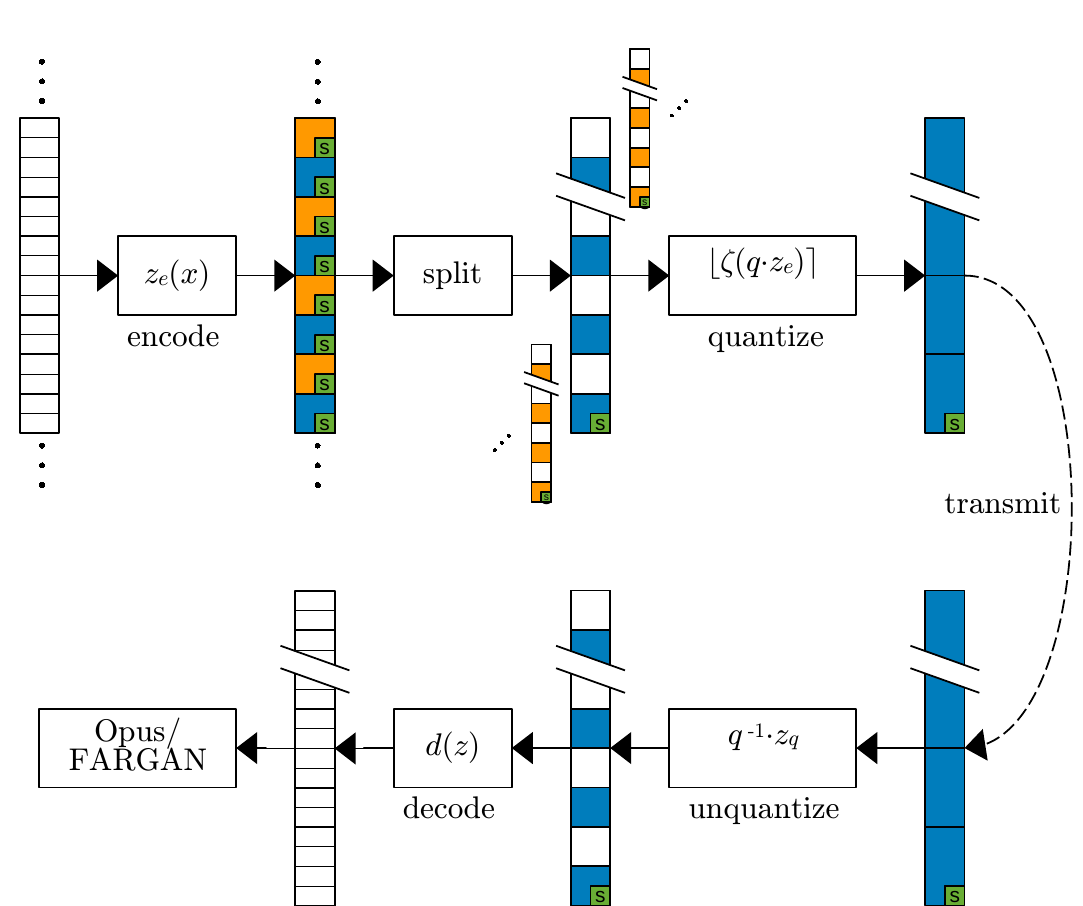}
\par\end{centering}
\caption{Encoding and decoding process. The encoder produces latent vectors
and initial states from acoustic features. The vectors are split into
overlapping redundancy packets and then quantized using a variable
resolution (the same vector can be quantized at different rates depending
on its position). At the receiver, the redundancy packets are entropy-decoded
and scaled back (unquantized) to recover the latent vectors. Those
are then decoded to produce 10\protect\nobreakdash-ms acoustic feature
vectors that can be used to synthesize audio in place of the missing
Opus packets. The redundancy decoding process happens only on-demand
such that no computation occurs when there is no loss.\protect\label{fig:RDO-VAE}}
\end{figure}

We propose a rate-distortion-optimized VAE (RDO-VAE) that directly
minimizes a rate-distortion loss function. From a sequence of input
vectors $\mathbf{x}\in\mathbb{R}^{L}$, the RDO-VAE produces an output
$\tilde{\mathbf{x}}\in\mathbb{R}^{L}$ by going through a sequence
of quantized latent vectors $\mathbf{z}_{q}\in\mathbb{Z}^{M}$, minimizing
the loss function
\begin{equation}
\mathcal{L}=D\left(\tilde{\mathbf{x}},\mathbf{x}\right)+\lambda H\left(\mathbf{z}_{q}\right)\ ,\label{eq:RDO}
\end{equation}
where $D\left(\cdot,\cdot\right)$ is the distortion loss, and $H\left(\cdot\right)$
denotes the entropy. The Lagrange multiplier $\lambda$ effectively
controls the target rate, with a higher value leading to a lower rate.
The high-level encoding and decoding process is illustrated in Fig.~\ref{fig:RDO-VAE}.

Because the latent vectors $\mathbf{z}_{q}$ are quantized, neither
$D\left(\cdot,\cdot\right)$ nor $H\left(\cdot\right)$ in~\eqref{eq:RDO}
are differentiable. For the distortion, a common way around the problem
is to use the \emph{straight-through} estimator~\cite{van2017neural,bengio2013estimating}.
More recently, various combinations involving ``soft'' quantization
-- through the addition of uniformly distributed noise -- have been
shown to produce better results~\cite{guo2021,balle2021}. In this
work, we choose to use a weighted average of the soft and straight-through
(hard quantization) distortions, as discussed in Sec.~\ref{subsec:Training}
and illustrated in Fig.~\ref{fig:training}.

\subsection{Rate estimator}

\begin{figure}
\begin{centering}
\includegraphics[width=0.9\columnwidth]{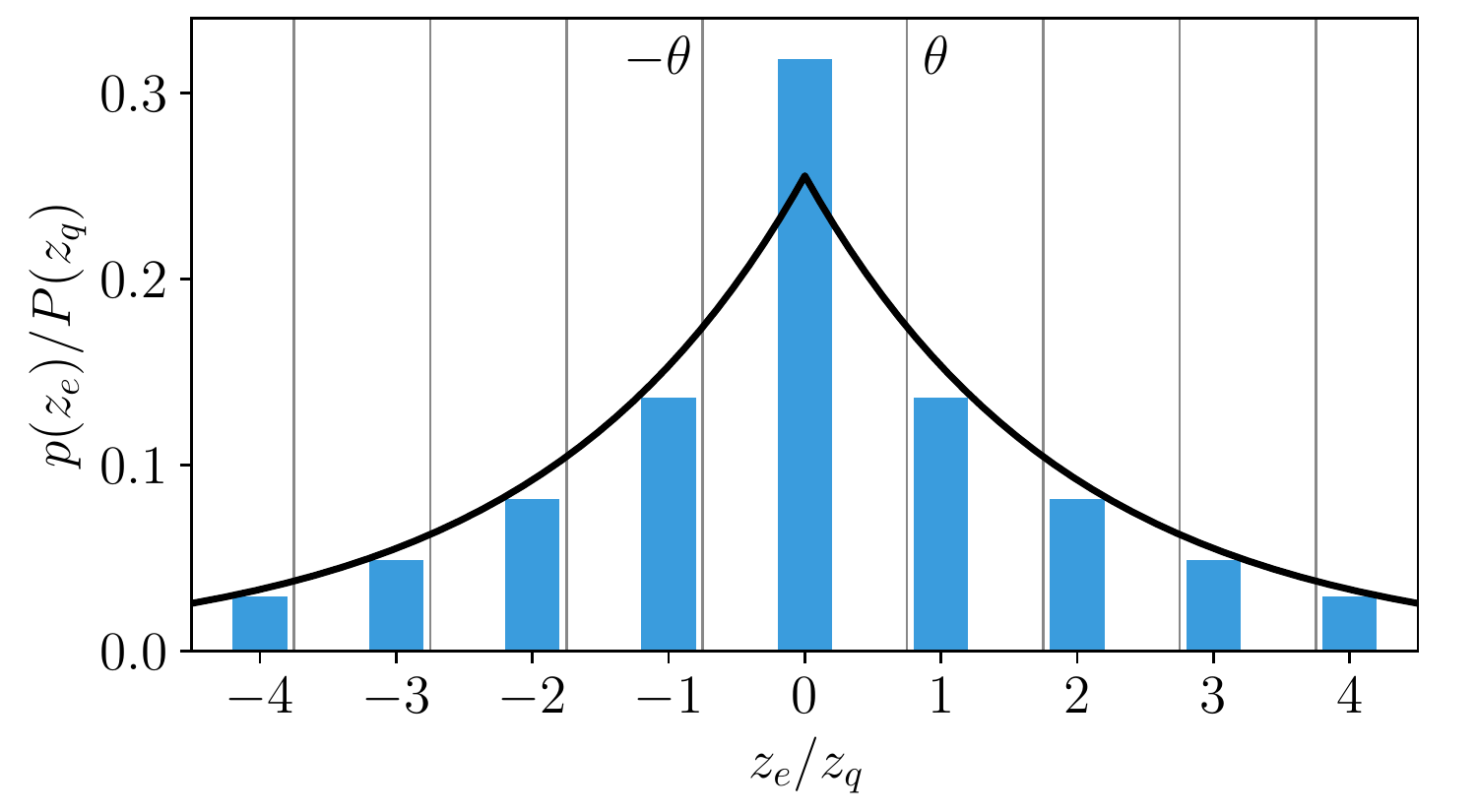}
\par\end{centering}
\caption{Continuous Laplace distribution $p\left(z_{e}\right)$ in black line
and the corresponding discrete distribution $P\left(z_{q}\right)$
in blue solid bars, for $r=0.6$ and $\theta=0.75$. The quantization
thresholds are shown as gray vertical lines. Note that the reconstruction
values are still integers and are not affected by $\theta$.\protect\label{fig:Laplace-distribution}}
\end{figure}

We use the Laplace distribution to model the latent space because
it is easy to manipulate and is relatively robust to probability modeling
mismatches. Since we can consider the rate of each variable independently,
let $z_{e}$ and $z_{q}$ represent one component of the unquantized
($\mathbf{z}_{e}$) and quantized ($\mathbf{z}_{q}$) vectors, respectively.
The continuous Laplace distribution is given by:
\begin{equation}
p\left(z_{e}\right)=-\frac{\log r}{2}r^{\left|z_{e}\right|}\ ,\label{eq:continuous_laplace}
\end{equation}
where $r$ is related to the standard deviation $\sigma$ by $r=e^{-\sqrt{2}/\sigma}$. 

An efficient way of quantizing a Laplace-distributed variable~\cite{sullivan1996}
is to use a fixed quantization step size, except around zero, where
all values of $q_{e}\in\left]-\theta,\theta\right[$ quantize to zero,
with $\theta>\frac{1}{2}$ arising from rate-distortion optimization.
We describe a quantizer with a step size of one without loss of generality,
since we can always scale the input and output to achieve the desired
quantization resolution. We thus define the quantizer as:
\begin{equation}
z_{q}=Q_{\theta}\left(z_{e}\right)=\mathrm{sgn}\left(z_{e}\right)\left\lfloor \max\left(\left|z_{e}\right|+1-\theta,\ 0\right)\right\rfloor \ ,\label{eq:quantization}
\end{equation}
where $\mathrm{sgn}\left(\cdot\right)$ denotes the sign function.
In the special case $\theta=1/2$, we simply round to the nearest
integer (ignoring ties).

We refer to the distribution of $z_{q}$ as the discrete Laplace distribution
(Fig.~\ref{fig:Laplace-distribution})
\begin{equation}
P\left(z_{q}\right)=\begin{cases}
1-r^{\theta} & z_{q}=0\\
\frac{1}{2}\left(1-r\right)r^{\left|z_{q}\right|+\theta-1} & z_{q}\neq0
\end{cases}\ .\label{eq:quantized_laplace}
\end{equation}
Since its entropy, $H\left(z_{q}\right)=\mathbb{E}\left[-\log_{2}P\left(z_{q}\right)\right]$,
is not differentiable with respect to $z_{e}$, we must find a way
to backpropagate the gradient. We find that using the straight-through
estimator for the rate results in very poor convergence, with the
training loss starting to increase again after a few epochs due to
the mismatch between the forward and backward pass of backpropagation. 

We seek to use a differentiable rate estimation on the unquantized
encoder output. An obvious choice is to use the differential entropy
$h\left(z_{e}\right)=\mathbb{E}\left[-\log_{2}p\left(z_{e}\right)\right]$,
which achieves better convergence. Unfortunately, the differential
entropy tends towards $-\infty$ when $p\left(z_{e}\right)$ becomes
degenerate as $r\rightarrow0$, which can cause many low-variance
latent variables to collapse to zero. Instead, we directly evaluate
the entropy of the discrete distribution $H\left(z_{e}\right)=\mathbb{E}\left[-\log_{2}P\left(z_{e}\right)\right]$
using the continuous latent $z_{e}$. To avoid the special case around
$z_{e}=0$, we use a value for $\theta$ for which both cases of~\eqref{eq:quantized_laplace}
are equal for $z_{e}=0$:
\begin{align}
1-r^{\theta} & =\frac{1}{2}\left(1-r\right)r^{\theta-1}\label{eq:no_special_case}\\
\theta & =\log_{r}\left(2r/\left(1+r\right)\right)\ .\label{eq:theta_implicit}
\end{align}
Using the value of $\theta$ given by~\eqref{eq:theta_implicit}
results in the generalized discrete entropy
\begin{equation}
H\left(z_{e}\right)=-\log_{2}\frac{1-r}{1+r}-\mathbb{E}\left[\left|z_{e}\right|\right]\log_{2}r\ .\label{eq:rate_loss}
\end{equation}
which has a form similar to an $L_{1}$ loss, but with a penalty term
$-\log_{2}\frac{1-r}{1+r}$ that counteracts the weighting factor
$\log_{2}r$. In the degenerate case where $r\rightarrow0$ (and thus
$z_{e}=0$), we have $H\left(z_{e}\right)=0$, which is the desired
behavior. An advantage of using~\eqref{eq:rate_loss} is that any
latent dimension that does not sufficiently reduce the distortion
to be ``worth'' its rate naturally becomes degenerate during training.
We can thus start with more latent dimensions than needed and let
the model decide on the number of useful dimensions. In practice,
we find that different values of $\lambda$ result in a different
number of non-degenerate pdfs (Fig.~\ref{fig:non-degenerate}).

\subsection{Quantization and encoding}

The dead zone, as defined by the quantizer $Q_{\theta}\left(z\right)$
in~\eqref{eq:quantization}, needs to be differentiable with respect
to both its input parameter $z$ and its width~$\theta$. That can
be achieved by implementing it as the differentiable function
\begin{equation}
\zeta\left(z\right)=z-\delta\tanh\frac{z}{\delta+\epsilon}\ ,\label{eq:dead-zone}
\end{equation}
where $\delta\approx\theta-1/2$ controls the width of the dead zone
and $\epsilon=0.1$ avoids training instabilities. The complete quantization
process thus becomes
\begin{equation}
z_{q}=\left\lfloor \zeta\left(q_{\lambda}\cdot z_{e}\right)\right\rceil \ ,\label{eq:diff_quantization}
\end{equation}
where $\left\lfloor \cdot\right\rceil $ denotes rounding to the nearest
integer, and $q_{\lambda}$ is the quantizer scale (higher $q_{\lambda}$
leads to higher quality). The quantizer scale $q_{\lambda}$ is learned
independently as an embedding matrix for each dimension of the latent
space and for each value of the rate-control parameter~$\lambda$. 

The quantized latent components $z_{q}$ can be entropy-coded~\cite{nigel1979}
using the discrete pdf in~\eqref{eq:quantized_laplace} parameterized
by $r$ and $\theta$. The value of $\theta$ is learned independently
of the quantizer dead-zone parameter $\delta$. Also, we learn a different
$r$ parameter for the soft and hard quantizers. The value of $\theta$
for the soft quantizer is implicit and thus does not need to be learned,
although a learned $\theta$ does not lead to significant rate reduction,
which is evidence that the implicit $\theta$ is close to the RD-optimal
choice. 

On the decoder side, the quantized latent vectors are entropy-decoded
and the scaling is undone:
\begin{equation}
z_{d}=q_{\lambda}^{-1}\cdot z_{q}\ .\label{eq:unquantize}
\end{equation}

Finally, we need to quantize the IS vector~$\mathbf{s}$ to be used
by the decoder. Although the encoder produces an IS at every frame,
only one IS per redundancy packet needs to be transmitted. In previous
work~\cite{valin2023dred}, we used a fixed-bitrate pyramid vector
quantizer (PVQ)~\cite{fischer1986pyramid} to transmit the IS vectors
since it made training simpler (only one rate). However, not being
able to optimize the bitrate as a function of $\lambda$ is sub-optimal
so we instead propose also using RDO-VAE for the IS. We can use the
same process as for encoding the latent vectors, with the exception
that the rate must be considered differently since only one IS is
transmitted for multiple latent vectors. Using the variable rate IS
allows the proposed redundancy mechanism to scale to a wider range
of bitrates than our previous work. 

\subsection{Encoder and decoder}

The encoder and decoder are constructed from a combination of gated
recurrent unit (GRU)~\cite{cho2014properties} and 1D convolutional
layers in time. To help with gradient propagation and avoid the vanishing
gradient problem, we introduce skip connections arranged like in the
DenseNet~\cite{huang2017densely} architecture. The encoder and decoder
networks each include 5~GRU layers alternating with 5~convolutional
layers. We found that the proposed DNN architecture with about 1~million
weights for each of the encoder and decoder achieves similar results
to the architecture of our previous work that used 2~million weights
for each (not considering the other improvements). 

\subsection{Training}

\label{subsec:Training}

During training, we vary $\lambda$ in such a way as to obtain average
rates between~8 and 80~bits per vector. We split the $\lambda$
range into 16~equally-spaced intervals in the log domain. For each
interval, we learn independent values for $q$, $\delta$, $\theta$,
as well as for the hard and soft versions of the Laplace parameter
$r$. To avoid giving too much weight to the low-bitrate cases because
of the large $\lambda$ values, we reduce the difference in losses
by weighting the total loss values by $1/\sqrt{\lambda}$:
\begin{equation}
\mathcal{L}=\frac{D\left(\tilde{\mathbf{x}},\mathbf{x}\right)}{\sqrt{\lambda}}+\sqrt{\lambda}\sum_{i=0}^{M-1}H\left(z_{e}^{(i)};r_{s}^{(i)}\right)\ .\label{eq:final_loss}
\end{equation}
The acoustic vector $\mathbf{x}=\left[\mathbf{c},p,v\right]$ includes
an 18-dimensional cepstrum $\mathbf{c}$, the log pitch frequency
$f_{0}$ and the voicing (pitch correlation) parameter $v$. The distortion
is then defined as 
\begin{equation}
D\left(\tilde{\mathbf{x}},\mathbf{x}\right)=E\left[\left\Vert \tilde{\mathbf{c}}-\mathbf{c}\right\Vert ^{2}+w_{p}\left|p-\tilde{p}\right|+\left|v-\tilde{v}\right|^{2}\right]\label{eq:distortion}
\end{equation}
where the pitch distortion weighting $w_{p}=10v^{2}$ ensures that
we only consider the pitch error for voiced signals. 

The large overlap between decoded sequences poses a challenge for
the training. Running a large number of overlapping decoders would
be computationally challenging. On the other hand, we find that decoding
the entire sequence produced by the encoder with a single decoder
leads to the model over-fitting to that particular case. We find that
encoding 4\nobreakdash-second sequences and randomly splitting them
into four non-overlapping sequences to be independently decoded leads
to acceptable performance and training time. From there, we can ensure
robustness to longer sequences by iteratively lenghtening the encoded
sequences, while maintaining a one-second average decoding duration.
The whole training setup is depicted in Fig.~\ref{fig:training}.

\input{training_figure.tex}

\section{Experiments \& Results}

\label{sec:Experiments-Results}

Both the RDO-VAE and the FARGAN vocoder are trained independently
on 205~hours of 16\nobreakdash-kHz speech from a combination of
TTS datasets~\cite{demirsahin-etal-2020-open,kjartansson-etal-2020-open,kjartansson-etal-tts-sltu2018,guevara-rukoz-etal-2020-crowdsourcing,he-etal-2020-open,oo-etal-2020-burmese,van-niekerk-etal-2017,gutkin-et-al-yoruba2020,bakhturina2021hi}
including more than 900~speakers in 34~languages and dialects. The
vocoder training is performed as described in~\cite{valin2024fargan}.

\begin{figure}

\begin{centering}
\includegraphics[width=0.85\columnwidth]{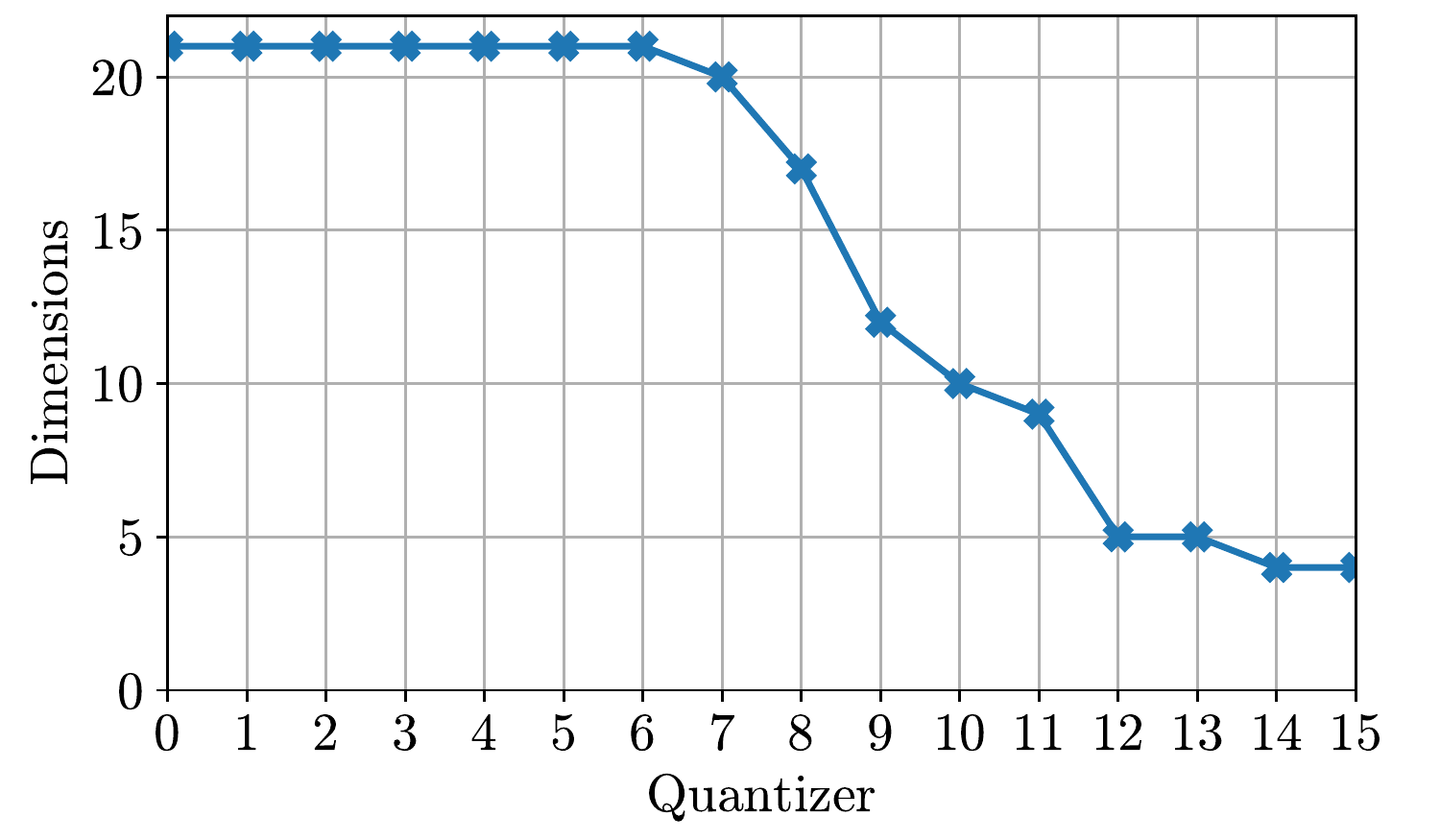}
\par\end{centering}
\caption{Number of non-degenerate dimensions for all 16~quantizers, where
quantizer 0 corresponds to the highest bitrate and quantizer 15 corresponds
to the lowest bitrate.\protect\label{fig:non-degenerate}}

\end{figure}

We train the RDO-VAE with $M=80$ initial latent dimensions and observe
that, depending on the quantizer, between 4~and 21~dimensions are
ultimately non-degenerate (Fig.~\ref{fig:non-degenerate}). At the
highest bitrate (quantizer 0), the latent vectors are coded with an
average of 72~bits each (equivalent to 1.8~kb/s), with 105~bits
per IS. At the lowest bitrate (quantizer 15), the latent vectors use
just 6~bits (equivalent to 150~b/s), with 48~bits per IS. 

We evaluate the proposed deep redundancy mechanism on speech compressed
with the Opus codec at 24~kb/s. We add 1.04~seconds of deep redundancy
in each 20-ms frame transmitted, so that 52~copies of every frame
are ultimately transmitted (concealing burst losses up to 1.02~seconds).
We vary the rate within each redundancy packet such that the average
rates are around 1250~b/s for the most recent frame and 150~b/s
for the oldest. The average rate over all frames is about 650~b/s,
including the IS, or 32~kb/s of total redundancy. 

A real-time C~implementation of DRED operating within the Opus codec
(including the FARGAN vocoder) is available under an open-source license\footnote{\href{https://gitlab.xiph.org/xiph/opus/-/tree/jstsp\_dred/dnn/torch/rdovae}{https://gitlab.xiph.org/xiph/opus/-/tree/jstsp\_dred/dnn/torch/rdovae}}.
The results correspond to the code released as part of Opus version~1.5\footnote{Demo samples provided at \href{https://www.opus-codec.org/demo/opus-1.5/}{https://www.opus-codec.org/demo/opus-1.5/}}. 

\subsection{Complexity}

The DRED encoder and decoder each have about 1~million~weights.
The encoder uses each weight once (multiply-add) for each 20\nobreakdash-ms
frame, resulting in a complexity of 100~MFLOPS. The decoder's complexity
varies depending on the loss pattern, but it can never decode more
than one latent vector every 40~ms on average. That results in a
worst-case average decoder complexity of 50~MFLOPS. Unlike the case
of the encoder, the decoder complexity can have bursts. On the receiver
side, the complexity is dominated by the FARGAN vocoder's 600~MFLOPS
worst-case complexity.

\subsection{Quality}

\label{subsec:Quality}

\begin{figure}

\begin{centering}
\includegraphics[width=0.85\columnwidth]{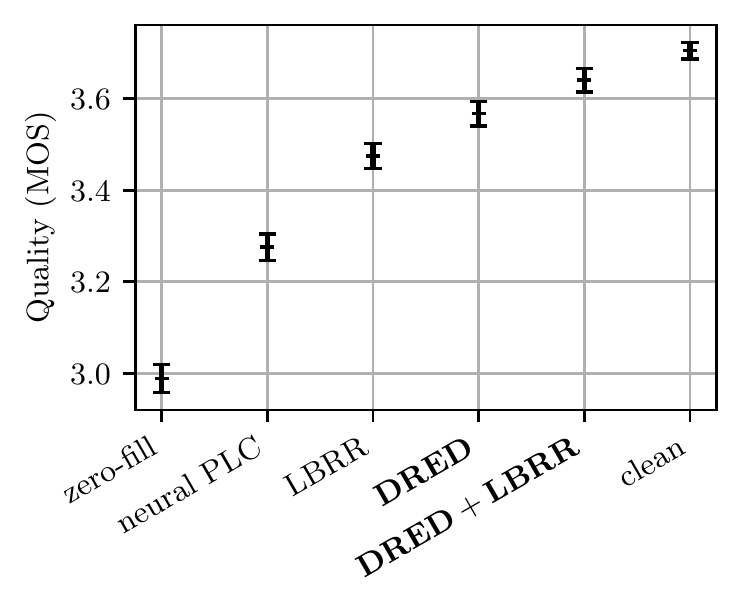}
\par\end{centering}
\caption{MOS results, including the 95\% confidence intervals. All differences
are statistically significant ($p<.05$).\protect\label{fig:MOS-evaluation}}

\end{figure}

We evaluated DRED on the PLC Challenge dataset~\cite{plc_challenge},
using the development test set files for both the audio and the recorded
packet loss sequences (18.4\%~average loss rate). The sequences have
losses ranging from 20~ms to bursts of up to one second, meaning
that the redundancy is able to cover all losses without the need for
regular PLC. We compare with the deep PLC results obtained in~\cite{valin2022plc}
(no redundancy), as well as with the original Opus LBRR\footnote{The total bitrate is increased to 40~kb/s to make room for LBRR,
which averages about 16~kb/s.}, both alone (requiring PLC) and combined with DRED (the first DRED
frame becomes unused). We also include an upper bound where DRED is
applied with uncompressed features. We include as anchors both clean/lossless
samples and samples where losses are replaced with zeros. 

The mean opinion score (MOS)~\cite{P.800} results in Fig.~\ref{fig:MOS-evaluation}
were obtained using the crowdsourcing methodology described in P.808~\cite{P.808,naderi2020open},
where each of the 966~test utterances was randomly assigned and evaluated
by 10~naive listeners (not all utterances were rated by the same
10~listeners). Listeners were asked to rate samples on an absolute
category rating scale from~1 (bad) to~5 (excellent). The results
show that DRED significantly outperforms both deep PLC and the existing
Opus LBRR. Despite the very low bitrate used for the redundancy, the
performance is already close to the uncompressed upper bound, suggesting
that the vocoder may already be the performance bottleneck. We also
note that LBRR and DRED appear to be complementary, with LBRR being
more efficient for short losses and DRED handling long losses.

\subsection{Complete system}

Since Opus is the mandatory-to-implement audio codec for the WebRTC
standard~\cite{rfc7874}, we evaluate DRED as part of a complete
real-time speech communications system. We modify the webrtc.org implementation
to add DRED support\footnote{\href{https://github.com/xiph/webrtc-opus-ng/tree/opus-ng}{https://github.com/xiph/webrtc-opus-ng/tree/opus-ng}}.
That evaluation includes not only the effect of the codec and redundancy,
but also the interaction with the jitter buffer. In practice when
losses occur, the jitter buffer increases the delay in a similar way
to the case where packets arrive later. From the jitter buffer's point
of view, a new DRED-containing packet arriving after a loss can be
viewed in the same way as the simultaneous arrival of all the lost
packets.

\begin{figure}
\begin{centering}
\includegraphics[width=1.05\columnwidth]{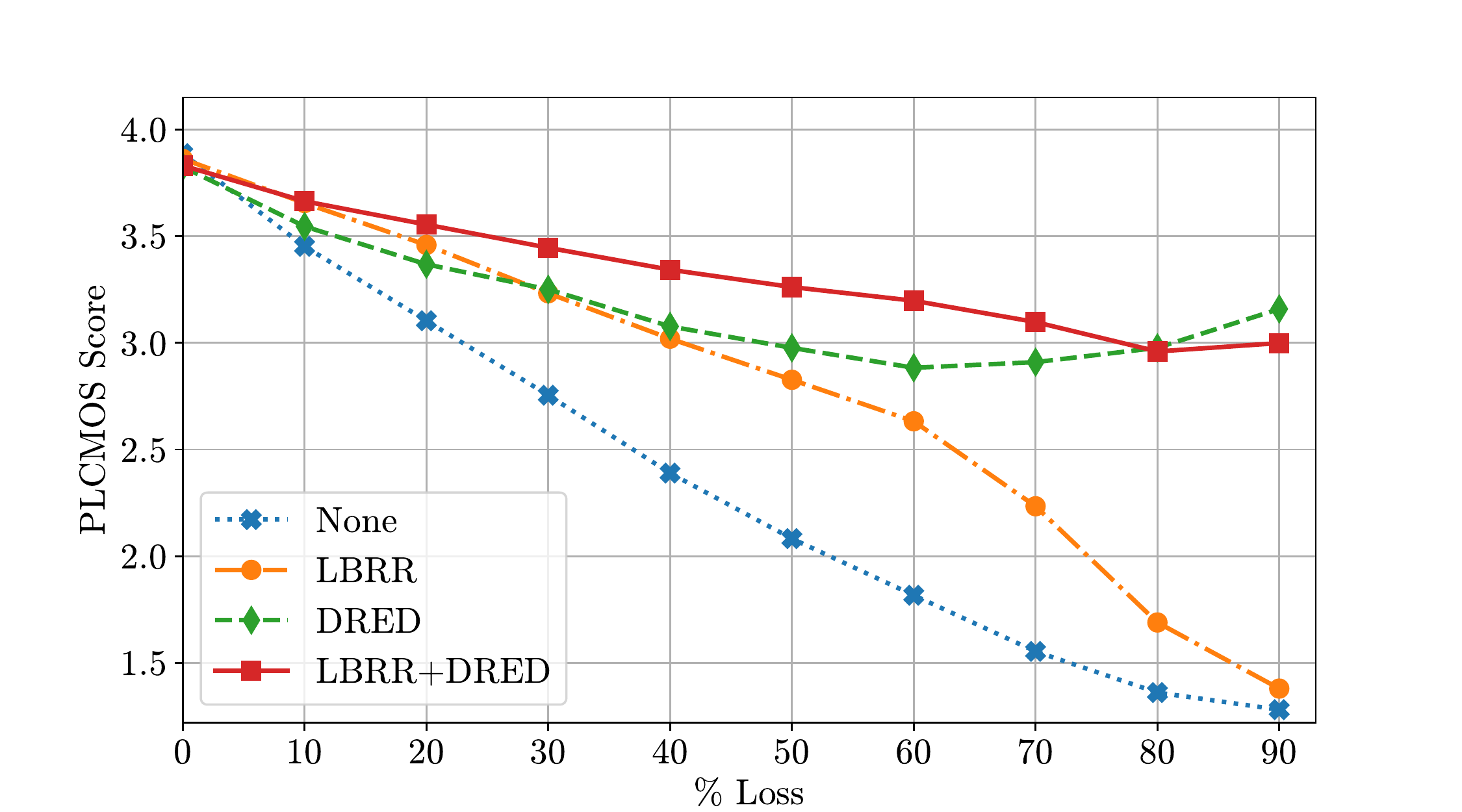}\caption{PLCMOSv2 objective evaluation of the different algorithms as a function
of the rate of packet loss.\protect\label{fig:PLCMOSv2-evaluation} }
\par\end{centering}
\end{figure}

For evaluation, we select a random subset of 200~samples from the
dataset used in Sec.~\ref{subsec:Quality}. We generate a single
packet loss pattern per-item using a generative model\footnote{\href{https://gitlab.xiph.org/xiph/opus/-/blob/main/dnn/lossgen.c}{https://gitlab.xiph.org/xiph/opus/-/blob/main/dnn/lossgen.c}}
trained from the real loss traces used in Sec.~\ref{subsec:Quality}
and conditioned on the average loss. We perform objective testing
using version~2 of the reference-free PLCMOS\footnote{\href{https://github.com/microsoft/PLC-Challenge/tree/main/PLCMOS}{https://github.com/microsoft/PLC-Challenge/tree/main/PLCMOS}}
algorithm~\cite{diener2023plcmos}. Results in Fig.~\ref{fig:PLCMOSv2-evaluation}
also demonstrate the effectiveness of DRED when compared to LBRR or
PLC alone.

\section{Conclusion}

We demonstrate the feasibility and benefits of transmitting large
amounts of low-bitrate audio redundancy to improve the robustness
of speech communication under burst packet loss. The proposed DRED
solution makes use of a rate-distortion-optimized VAE operating in
a forward-backward configuration to compute and quantize a sequence
of latent vectors at a 40\nobreakdash-ms interval and transmit self-contained
overlapping segments of redundancy to the receiver. We not only show
that DRED is more effective than the Opus LBRR redundancy for high
loss rates, but that the two redundancy methods are complementary.
Taking advantage of the proposed DRED requires an adaptive jitter
buffer. Even though the optimal trade-offs between loss robustness
and jitter buffer delay is still an open question, our results in
the context of WebRTC demonstrates the practical effectiveness of
DRED. The proposed deep redundancy is currently being standardized
at the IETF~\cite{valin2024ietfdred} as an extension of the Opus
codec.

\balance

\bibliographystyle{IEEEbib}
\bibliography{vae,plc,corpora,lpcnet}

\end{document}

%% file: training_figure.tex
\tikzstyle{arrow}=[draw, -latex]

\definecolor{squidink}{RGB}{35, 47, 62}
\definecolor{anchor}{RGB}{0, 49, 129}
\definecolor{sky}{RGB}{32, 116, 213}
\definecolor{rind}{RGB}{251, 216, 191}
\definecolor{smile}{RGB}{255, 153, 0}

\begin{figure}
\begin{center}
\scalebox{0.7}{
\begin{tikzpicture}[x=10pt, y=10pt]

\tikzstyle{quantizer} = [rectangle, minimum width=3, minimum height=15 , text centered, fill=anchor!40]

\tikzstyle{scaler} = [rectangle, minimum width=30, minimum height=15 , rounded corners, text centered, fill=anchor!40]

\tikzstyle{slicer} = [rectangle, minimum width=30, minimum height=15 , rounded corners, text centered, fill=anchor!20]

\tikzstyle{encdec} = [trapezium, trapezium angle=-65, minimum width=8, minimum height=20, text centered, fill=sky!30!smile!30!rind!30!anchor!70, rotate=90]

\node [encdec] (enc) at (-2, 0.25) {Enc};
\node [scaler] (scale) at (3, 1) {scale};
\node [scaler] (scale2) at (3, -0.5) {scale};
\node [slicer] (slicer) at (10, 1) {slice};
\node [slicer] (selecter) at (10, -0.5) {select};

\node [quantizer] (softquant) at (14.5, -5) {soft Q};
\node [quantizer] (hardquant) at (19.5, -5) {hard Q};

\node [scaler] (invscale1) at (3, -10) {$\text{scale}^{-1}$};
\node [scaler] (invscale2) at (3, -17) {$\text{scale}^{-1}$};
\node [scaler] (invscale3) at (3, -11.5) {$\text{scale}^{-1}$};
\node [scaler] (invscale4) at (3, -18.5) {$\text{scale}^{-1}$};

\node [encdec] (sdec) at (-2, -10.75) {Dec};
\node [encdec] (hdec) at (-2, -17.75) {Dec};

\tikzstyle{featurearrow} = [arrow, line width=2, draw=sky!30];
\tikzstyle{latentarrow} = [arrow, line width=2, draw=smile!30];
\tikzstyle{statearrow} = [arrow, line width=2, draw=rind!50];

\draw [latentarrow] (-1, 1) -- (scale.west);

\draw [latentarrow] (scale) -- (slicer);
\draw [latentarrow] (slicer) -- (15, 1) -- (15, -4.25);
\draw [latentarrow] (15, 1) -- (20, 1) -- (20, -4.25);
\draw [latentarrow] (15, -5.75) -- (15, -10) -- (invscale1);
\draw [latentarrow] (20, -5.75) -- (20, -17) -- (invscale2);
\draw [latentarrow] (invscale1) -- (-1, -10);
\draw [latentarrow] (invscale2) -- (-1, -17);

\draw [statearrow] (-1, -0.5) -- (scale2.west);
\draw [statearrow] (scale2.east) -- (selecter.west);
\draw [statearrow] (selecter.east) -- (14, -0.5) -- (14, -4.25);
\draw [statearrow] (14, -0.5) -- (19, -0.5) -- (19, -4.25);
\draw [statearrow] (14, -5.75) -- (14, -11.5) -- (invscale3);
\draw [statearrow] (invscale3) -- (-1, -11.5);
\draw [statearrow] (19, -5.75) -- (19, -18.5) -- (invscale4);
\draw [statearrow] (invscale4) -- (-1, -18.5);

\draw [featurearrow] (-14, 0.25) -- (enc);
\draw [featurearrow] (sdec) -- (-14, -10.75);
\draw [featurearrow] (hdec) -- (-14, -17.75);

\node [circle, fill=smile!60, minimum size=4.5] (rate) at (7, -5) {\small$\sqrt\lambda H(z, p_{\lambda})$};
\node [circle, fill=smile!60, minimum size=4.5] (harddist) at (-10, -3.5) {$\frac{D(x, \hat x_h)}{\sqrt{\lambda}}$};
\node [circle, fill=smile!60, minimum size=4.5] (softdist) at (-6, -7) {$\frac{D(x, \hat x_s)}{\sqrt\lambda}$};

\draw [arrow, dashed] (rate) -- (6, 1);
\draw [arrow, dashed] (rate) -- (7, -0.5);

\draw [arrow, dashed] (harddist) -- (-10, 0.05);
\draw [arrow, dashed] (harddist) -- (-10, -17.55);

\draw [arrow, dashed] (softdist) -- (-6, 0.05);
\draw [arrow, dashed] (softdist) -- (-6, -10.55);

\node (lambda1) at (17.5, -10) {$\lambda$};
\draw [arrow, dashed] (lambda1) -- (softquant);
\draw [arrow, dashed] (lambda1) -- (hardquant);

\node (lambda2) at (3, -14) {$\lambda$};
\draw [arrow, dashed] (lambda2) -- (invscale3);
\draw [arrow, dashed] (lambda2) -- (invscale2);

\node (lambda3) at (3, -5) {$\lambda$};
\draw [arrow, dashed] (lambda3) -- (scale2);

\def\xoff{5}

\draw [featurearrow] (-10 + \xoff, -22) -- (-8 + \xoff, -22);
\node (featurelegend) at (-9 + \xoff, -23) {features};

\draw [latentarrow] (-3 + \xoff, -22) -- (-1 + \xoff, -22);
\node (letentlegend) at (-2 + \xoff, -23) {latent vectors};

\draw [statearrow] (4 + \xoff, -22) -- (6 + \xoff, -22);
\node (initstatelegend) at (5 + \xoff, -23) {initial states};

\end{tikzpicture}
}
\caption{Setup for training RDO-VAE. Features are encoded by the encoder to produce a sequence of latent vectors and initial states. In \textit{slice} a sub-sequence is sliced from the sequence of latent vectors and in \textit{select} the matching initial state is selected. Initial state and latent vectors are passed both through a hard quantization and a soft quantization unit and the outputs of both units are decoded separately.}
\label{fig:training}
\end{center}
\end{figure}